\documentclass[11pt]{article}
\usepackage{moriond,epsfig}
\usepackage{epstopdf}
\pdfoutput=1

\def\Journal#1#2#3#4{{#1} {\bf #2}, #3 (#4)}

\def\NIMA{{\em Nucl. Instrum. Methods} A}
\def\NPB{{\em Nucl. Phys.} B}

\def\PRL{\em Phys. Rev. Lett.}
\def\PRD{{\em Phys. Rev.} D}

\def\ANP{\em Adv. Nucl. Phys.}

\def\AIP{\em AIP Conf. Proc.}

\def\be{\begin{equation}}
\def\ee{\end{equation}}
\def\bea{\begin{eqnarray}}
\def\eea{\end{eqnarray}}

\begin{document}
\vspace*{4cm}
\title{$W$ Boson Production in Polarized p+p Collisions at RHIC}

\author{ J.R. Stevens for the STAR Collaboration }

\address{Department of Physics and CEEM, Indiana University, Bloomington, IN 47408 USA} 

\maketitle\abstracts{
The production of $W^{\pm}$ bosons in longitudinally polarized $\vec{p}+\vec{p}$ collisions at RHIC provides a new means of studying the spin-flavor asymmetries of the proton sea quark distributions.  Details of the $W^{\pm}$ event selection in the $e^{\pm}$ decay channel at mid-rapidity are presented, along with preliminary results for the production cross section and parity-violating single-spin asymmetry, $A_L$, from the STAR Collaboration's 2009 data at $\sqrt{s}=500$ GeV.}

\section{Introduction}
The high-energy polarized $\vec{p}+\vec{p}$ collisions at $\sqrt{s}=500$ GeV at RHIC provide a unique way to study one of the unresolved questions in Quantum Chromodynamics (QCD), namely the spin structure of the nucleon.  Polarized deep-inelastic scattering (DIS) experiments have shown that the contribution of quark spins to that of the proton is surprisingly small, at the level of $\sim 25\%$.~\cite{dis}  Inclusive DIS measurements sum over quark flavor and are only sensitive to the combined contributions of quarks and anti-quarks.  Semi-inclusive DIS measurements, however, can achieve separation of the quark and anti-quark spin contributions by flavor and have been included, along with RHIC data constraining the gluon spin contribution, in a recent global analysis.~\cite{dssv}  The extracted anti-quark polarized Parton Distribution Functions (PDFs) have sizable uncertainties compared to the well-constrained quark + anti-quark sums.  

$W^{-(+)}$ bosons are produced at leading order through $\bar{u}+d\,(\bar{d}+u)$ interactions at the partonic level, and can be detected through their leptonic decays.  The parity-violating nature of the weak production process gives rise to large longitudinal single-spin asymmetries, $A_L$, which yield a direct and independent probe of the quark and anti-quark polarized PDFs.  The asymmetry is defined as $A_L=(\sigma^+ - \sigma^-)/(\sigma^+ + \sigma^-)$, where $\sigma^{+(-)}$ refer to the cross section when the helicity of the polarized proton beam is positive (negative).  Theoretical frameworks have been developed to describe the production of $W$ bosons and their decay leptons in polarized $\vec{p}+\vec{p}$ collisions.~\cite{rhicbos,nlo}   

\section{Experimental Details}

The STAR~\cite{star} detector systems used in this measurement are the Time Projection Chamber (TPC), which provides tracking of charged particles in a 0.5 T solenoidal magnetic field for pseudorapidities of $|\eta|<1.3$, and the Barrel and Endcap Electromagnetic Calorimeters (BEMC, EEMC), which are lead-scintillator sampling calorimeters covering $|\eta|<1$ and $1.09<\eta<2$, respectively.  Each of these detectors provides $2\pi$ coverage in the azimuthal angle, $\phi$.  

The data presented in this contribution were accumulated in 2009 when, for the first time, a significant dataset was collected for polarized proton collisions at a center of mass energy of $\sqrt{s}=500$ GeV.  The beam polarizations, measured with Coulomb-Nuclear Interference (CNI) proton-carbon polarimeters~\cite{cni} and calibrated with a polarized hydrogen gas-jet target,~\cite{pH} averaged $(38 \pm 8.3)\%$ and $(40 \pm 12.1)\%$ for the two beams.  The trigger condition used to select events online for this measurment involved a two-stage energy requirement in the BEMC.  Electrons and positrons from $W$ decays at mid-rapidity are characterized by a large transverse energy, $E_T$, peaked at $\sim M_{W}/2$ (Jacobian peak).  To pass the trigger, an event must satisfy a hardware threshold of $E_T > 7.3$ GeV in a single BEMC tower.  Additionally a software level trigger searched for a seed tower with $E_T > 5$ GeV and required that the maximum $2 \times 2$ tower cluster including that seed have an $E_T$ sum larger than 13 GeV.  The cross section for the hardware level trigger was measured using the vernier scan technique~\cite{vernier} to be 481 $\pm$ 10 (stat.) $\pm$ 110 (syst.) nb, which yields an integrated luminosity of 13.7 $\pm$ 0.3 (stat.) $\pm$ 3.1 (syst.) pb$^{-1}$ for this data set.  The systematic uncertainty in the luminosity is dominated by possible non-gaussian components of the beam profile seen in the vernier scan data. 

\begin{figure}
  \begin{center}
    \includegraphics[width=\textwidth]{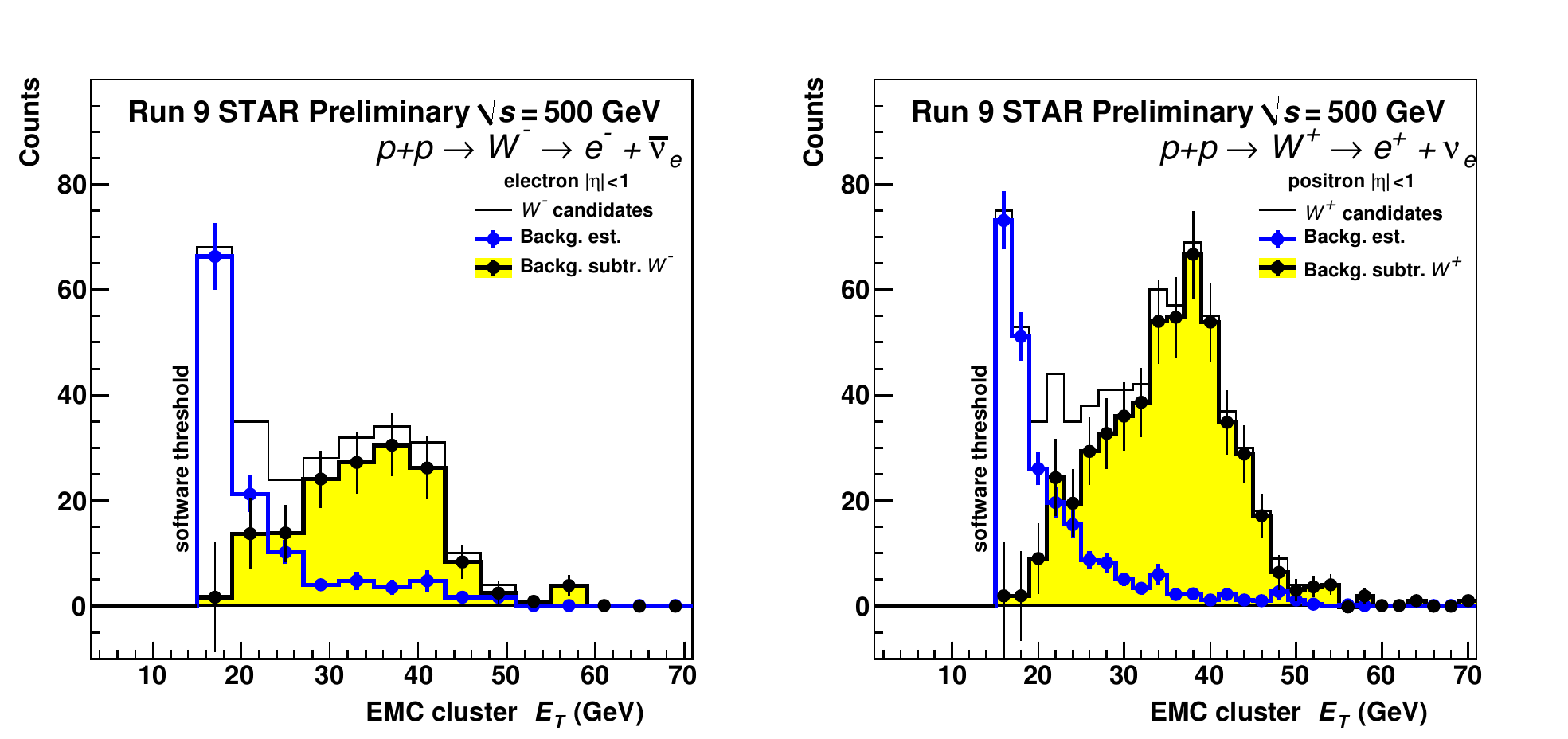}
    \caption{Candidate $e^{\pm}$ BEMC cluster $E_T$ for $W^-$ (left) and $W^+$ (right).}
    \label{fig:jacob}
  \end{center}
\end{figure}

$W$ candidate events were selected offline based on kinematical and topological differences between leptonic $W^{\pm}$ decay, and QCD background events.  $W^{\pm} \rightarrow e^{\pm}+\nu$ decay events contain a nearly isolated $e^{\pm}$ and an undetected neutrino opposite in azimuth, leading to a large missing $E_T$, characterized by a large imbalance in the vector $p_T$ sum of all reconstructed final-state objects.  Background QCD (e.g., di-jet) events, however, typically have a small imbalance in the vector $p_T$ sum.  An $e^{\pm}$ candidate is thus defined to be any TPC track with $p_T > 10$ GeV which originated from a primary vertex with $|z| < 100$ cm, where $z$ is the direction along the beamline from the nominal interaction point.  The candidate track is required to point to a $2 \times 2$ BEMC tower cluster with an $E_T$ sum, $E_T^{e}$, larger than $15$ GeV and whose centroid is less than $7$ cm from the extrapolated track.  Furthermore, two isolation cuts are imposed on the candidate: a) a requirement that the excess $E_T$ in the surrounding $4 \times 4$ tower cluster be less than $5\%$ of $E_T^{e}$, and b) that the excess EMC tower + TPC track $E_T$ sum within a cone radius $R=0.7$ of the cadidate in $\eta$-$\phi$ space be less than $12\%$ of $E_T^{e}$.  The final two selection criteria are based on cuts on the `away-side' $E_T$ and the vector $p_T$ sum, whose meaning and construction are as follows.  The away-side $E_T$ is the EMC + TPC $E_T$ sum over the full $\eta$ range and $\phi \in [\phi_e + \pi + 0.7,\phi_e + \pi - 0.7]$.  The vector $p_T$ sum is defined as the sum of the $e^{\pm}$ candidate $p_T$ vector and the $p_T$ vectors of all the reconstructed jets with thrust axes outside the $R=0.7$ cone around the candidate.  Jets were reconstructed using the standard mid-point cone algorithm used in previous STAR jet measurements.~\cite{jet}  The final $W$ candidate events are selected by requiring the away-side $E_T$ to be less than 30 GeV and the magnitude of the vector $p_T$ sum to be larger than 15 GeV.  

Figure~\ref{fig:jacob} shows the charge-separated yield as a function of the candidate $2 \times 2$ BEMC tower cluster $E_T$, or $E_T^{e}$, for events satisfying the selection criteria above.  The $W$ candidates show the characteristic Jacobian peak at $E_T^{e} \sim M_{W}/2$.  The efficiency for reconstructing $W^{\pm} \rightarrow e^{\pm}+\nu$ events was evaluated using PYTHIA Monte-Carlo with full GEANT simulation of the detector response.  Within the imposed kinematic acceptance of the decay $e^{\pm}$ , $|\eta_{e}|<1$ and $E_T^{e}>25$ GeV, the overall reconstruction efficiency was estimated to be 56$\%$.

\begin{figure}
  \begin{center}
    \includegraphics[height=98mm]{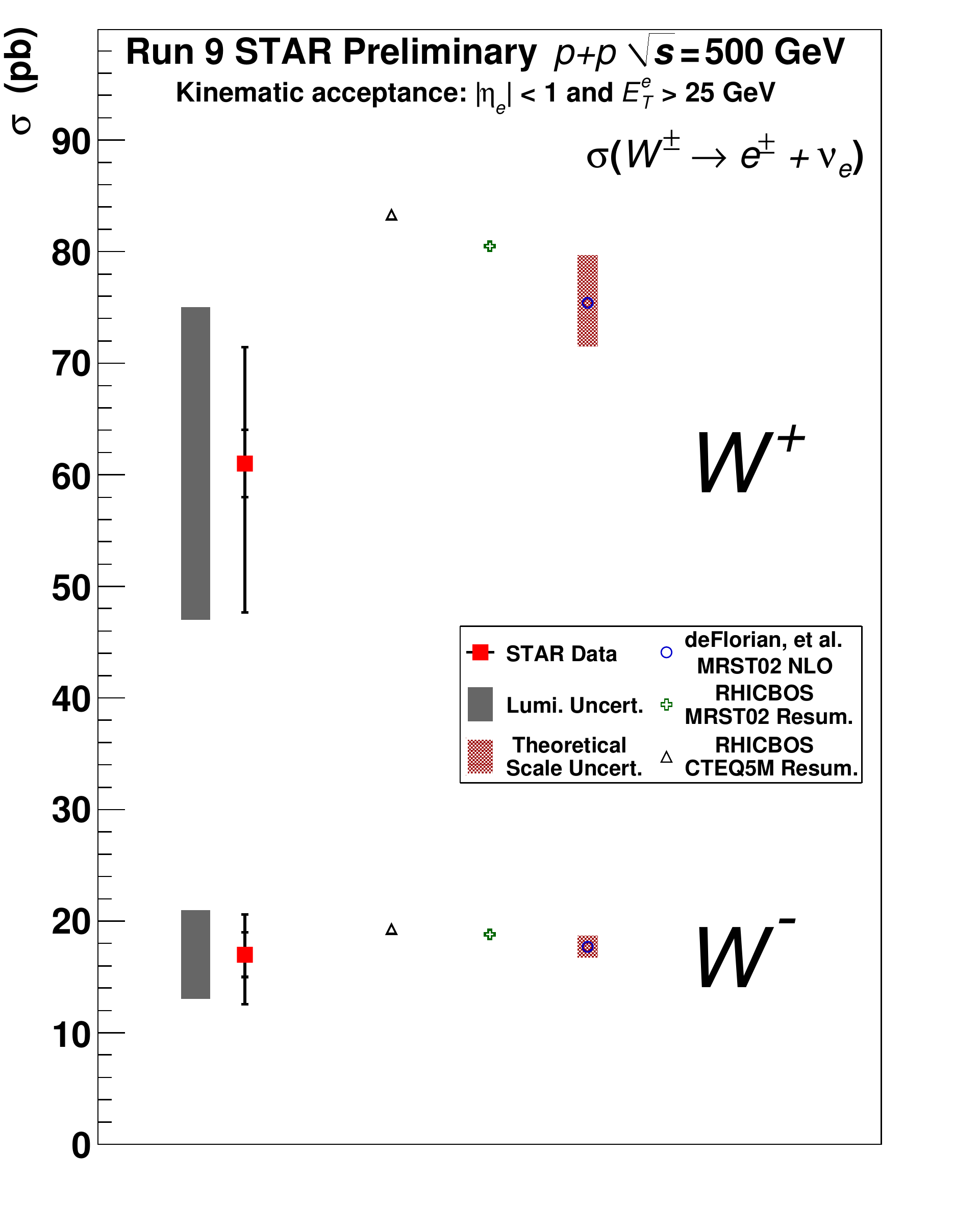}
    \includegraphics[height=107mm]{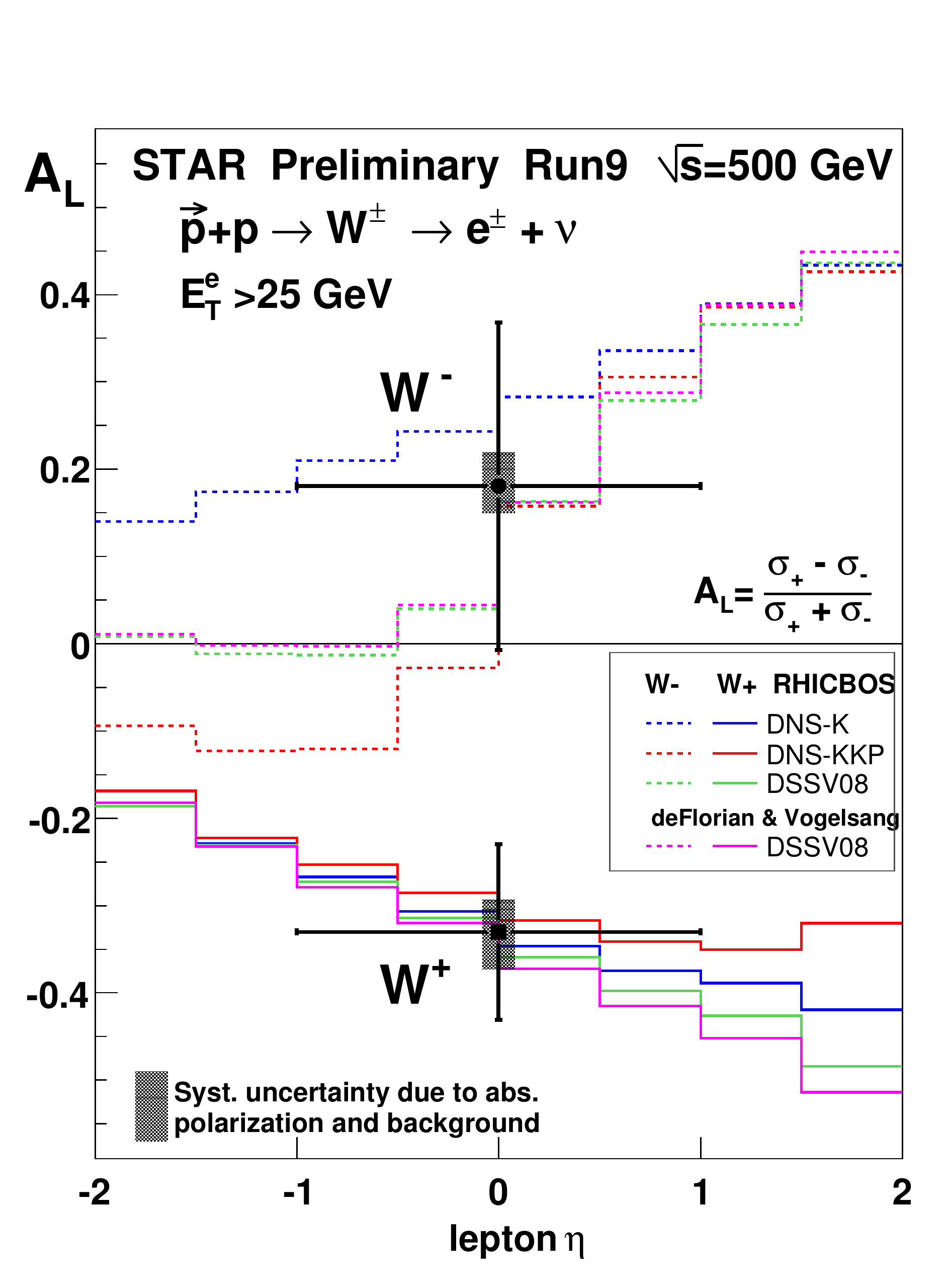}
    \caption{Measured cross section (left) and longitudinal single-spin asymmetry, $A_L$, (right) for $W^{\pm}$ production.}
    \label{fig:xsec}
  \end{center}
\end{figure}

The amount of background remaining in the $W$ candidate sample, after applying the selection criteria described above, is estimated from three contributions.  The first contribution is from $W^{\pm} \rightarrow \tau^{\pm} + \nu$ decay where the $\tau^{\pm}$ decays semi-leptonically to an $e^{\pm}$.  This background was estimated using a similar Monte-Carlo simulation as was used for the efficiency.  Another contribution estimates the impact of the missing calorimetric coverage for $-2<\eta<-1.09$.  To determine this contribution to the background, the analysis was performed a second time with the EEMC not used as an active detector.  The difference in the $W$ candidate $E_T^{e}$ distribution with and without the EEMC included in the analysis is taken to be the estimate for the missing calorimetric coverage.  The final contribution is estimated by normalizing a data-driven background shape to the remaining $W$ candidate signal, in the $E_T^{e}$ range below 19 GeV, after the first two background contributions are removed.  This data-driven background shape is obtained by inverting the last two requirements in the $W$ candidate selection, namely requiring that the away-side $E_T$ be greater than 30 GeV or the magnitude of the vector $p_T$ sum be less than 15 GeV.  The total background and the remaining background-subtracted spectra are shown in Figure~\ref{fig:jacob} in the blue and yellow histograms, respectively.  A systematic uncertainty for the background estimation was determined by varying the inverted cuts used to obtain the data-driven background shape and by varying the range where the background shape is normalized to the remaining $W$ candidate signal.   

\section{Results}


Preliminary results for the production cross section of $W^{\pm} \rightarrow e^{\pm}+\nu$ from candidate events with $|\eta_{e}|<1$ and $E_T^{e}>25$ GeV are shown in Figure~\ref{fig:xsec} (left).  The measured values are $\sigma(W^{+} \rightarrow e^{+}+\nu) =$ 61 $\pm$ 3 (stat.) $^{+10}_{-13}$ (syst.) $\pm$ 14 (lumi.) pb and $\sigma(W^{-} \rightarrow e^{-}+\bar{\nu}) =$  17 $\pm$ 2 (stat.) $^{+3}_{-4}$ (syst.) $\pm$ 4 (lumi.) pb.  The statistical and systematic uncertainties are shown as the error bars on the red data points.  The systematic uncertainty of the measured luminosity, shown separately as the grey bands in Figure~\ref{fig:xsec} (left), is dominated by the uncertainty in the vernier scan measurement mentioned previously.  The measured cross sections are consistent with predictions based on full resummed~\cite{rhicbos} and NLO~\cite{nlo} calculations, which are also shown in Figure~\ref{fig:xsec} (left).

Figure~\ref{fig:xsec} (right) presents preliminary results for the single-spin asymmetry $A_L$ for candidate events with $|\eta_{e}|<1$ and $E_T^{e}>25$ GeV.  The measured asymmetries are $A_L^{W^{+}}=$ -0.33 $\pm$ 0.10 (stat.) $\pm$ 0.04  (syst.) and $A_L^{W^{-}}=$ 0.18 $\pm$ 0.19 (stat.) $^{+0.04}_{-0.03}$ (syst.).  The systematic uncertainties, shown as the grey bands in figure~\ref{fig:xsec} (right), are dominated by the absolute uncertainties of the beam polarization.  The four curves shown are predictions based on full resummed~\cite{rhicbos} and NLO~\cite{nlo} calculations for two polarized PDFs,~\cite{dssv,dns} which are constrained by inclusive and semi-inclusive DIS measurements.  Our results are consistent with the theoretical expectations. 

\section{Summary and Outlook}
The study of parity-violating single-spin asymmetries for $W^{\pm}$ bosons produced in polarized $\vec{p}+\vec{p}$ collisions offers a clean and unique approach to probe the flavor and spin structure of the proton.  Presented in this contribution are the first measurements of $W^{\pm}$ single-spin asymmetries, $A_L$, and production cross sections by the STAR Collaboration, both of which are consistent with resummed and NLO calculations.  Future planned STAR measurements at mid-rapidity and forward rapidity with increased luminosity and beam polarization are expected to provide significant constraints on the polarized QCD sea.

\end{document}